%% file: main.tex
\documentclass[sigconf]{acmart}

\usepackage{censor}
\input{_packages}

\input{_commands}
\begin{document}

\title[A Grassroots Network and Community Roadmap for Interconnected Autonomous Science Laboratories]{
A Grassroots Network and Community Roadmap\\for Interconnected Autonomous Science Laboratories\\for Accelerated Discovery
}

\input{_authors}

\thanks{Notice: This manuscript has been authored by UT-Battelle, LLC, under contract DE-AC05-00OR22725 with the US Department of Energy (DOE). The publisher acknowledges the US government license to provide public access under the DOE Public Access Plan (\url{http://energy.gov/downloads/doe-public-access-plan}).}

\input{abstract}

\maketitle

\input{sec_introduction}
\input{sec_aisle}
\input{sec_dimensions}
\input{sec_conclusion}

{
\small
\pp{Acknowledgments}
This research used resources of the Oak Ridge Leadership Computing Facility, and is sponsored by the INTERSECT Initiative as part of the Laboratory Directed Research and Development Program of Oak Ridge National Laboratory, supported by the Office of Science of the U.S. Department of Energy under Contract No. DE-AC05-00OR22725.
}

\bibliographystyle{ACM-Reference-Format}

\end{document}

%% file: _packages.tex
\usepackage{amsmath,amsfonts}
\usepackage{algorithmic}
\usepackage{hyperref}
\usepackage{graphicx}
\usepackage{paralist}
\usepackage{tcolorbox}
\usepackage{textcomp}
\usepackage{xcolor}
\usepackage{xspace}
\usepackage{enumitem}

%% file: _commands.tex
\definecolor{background}{HTML}{F4F0EB}
\definecolor{ruler}{HTML}{8E8D8C}

\newcommand{\milestones}[2][inline]{
\begin{tcolorbox}[
    width=\linewidth,
    colback={background},
    colframe={ruler},
    boxsep=2pt,          
    top=1.5pt,             
    bottom=1.5pt,           
    left=4pt,
    right=4pt
 ]
 \textbf{\color{ruler} \footnotesize MILESTONES:} \small #2 \xspace \end{tcolorbox}}

\newcommand{\pp}[2][inline]{
\smallskip\noindent\textbf{\emph{#2.}}
}
\setcopyright{acmlicensed}
\copyrightyear{}
\acmYear{}
\acmDOI{}

\acmConference[]{}{}{}
\acmISBN{}

%% file: _authors.tex
\settopmatter{authorsperrow=4} 

\author{Rafael Ferreira da Silva}
\orcid{0000-0002-1720-0928}
\affiliation{
  \institution{Oak Ridge Nat. Lab.}
  \city{Oak Ridge, TN}
  \country{USA}
}

\author{Milad Abolhasani}
\orcid{0000-0002-8863-3085}
\affiliation{
  \institution{North Carolina State University}
  \city{Raleigh, NC}
  \country{USA}
}
\author{Dionysios A. Antonopoulos}
\affiliation{
  \institution{Argonne Nat. Lab.}
  \city{Lemont, IL}
  \country{USA}
}
\author{Laura Biven}
\orcid{0000-0002-5755-8449}
\affiliation{
  \institution{Jefferson Lab.}
  \city{Newport News, VA}
  \country{USA}
}
\author{Ryan Coffee}
\orcid{0000-0002-2619-8823}
\affiliation{
  \institution{SLAC Nat. Accelerator Lab.}
  \city{ Menlo Park, CA}
  \country{USA}
}
\author{Ian T. Foster}
\orcid{0000-0003-2129-5269}
\affiliation{
  \institution{Argonne Nat. Lab.}
  \city{Lemont, IL}
  \country{USA}
}
\affiliation{
  \institution{University of Chicago}
  \city{Chicago, IL}
  \country{USA}
}
\author{Leslie Hamilton}
\affiliation{
  \institution{Johns Hopkins University Applied Physics Laboratory}
  \city{Laurel, MD}
  \country{USA}
}
\author{Shantenu Jha}
\orcid{0000-0002-5040-026X}
\affiliation{
  \institution{Rutgers, New Brunswick \\ PPPL \&  Princeton Univ.}
  \city{Princeton, NJ}
  \country{USA}
}
\author{Theresa Mayer}
\orcid{0009-0000-2105-2529}
\affiliation{
  \institution{Carnegie Mellon University}
  \city{Pittsburgh, PA}
  \country{USA}
}
\author{Benjamin Mintz}
\orcid{0000-0002-4054-1229}
\affiliation{
  \institution{Oak Ridge Nat. Lab.}
  \city{Oak Ridge, TN}
  \country{USA}
}
\author{Robert G. Moore}
\affiliation{
  \institution{Oak Ridge Nat. Lab.}
  \city{Oak Ridge, TN}
  \country{USA}
}
\author{Salahudin Nimer}
\orcid{0000-0001-7298-3358}
\affiliation{
  \institution{Johns Hopkins University Applied Physics Laboratory}
  \city{Laurel, MD}
  \country{USA}
}
\author{Noah Paulson}
\affiliation{
  \institution{Argonne Nat. Lab.}
  \city{Lemont, IL}
  \country{USA}
}

\author{Woong Shin}
\orcid{0000-0001-7207-7814}
\affiliation{
  \institution{Oak Ridge Nat. Lab.}
  \city{Oak Ridge, TN}
  \country{USA}
}
\author{Frédéric Suter}
\orcid{0000-0003-1902-1955}
\affiliation{
  \institution{Oak Ridge Nat. Lab.}
  \city{Oak Ridge, TN}
  \country{USA}
}
\author{Mitra Taheri}
\affiliation{
  \institution{Johns Hopkins University}
  \city{Baltimore, MD}
  \country{USA}\\
  \institution{Pacific Northwest Nat. Lab.}
  \city{Richland, WA}
  \country{USA}
}
\author{Michela Taufer}
\orcid{0000-0002-0031-6377}
\affiliation{
  \institution{University of Tennessee}
  \city{Knoxville, TN}
  \country{USA}
}
\author{Newell R. Washburn}
\orcid{0000-0001-7843-8860}
\affiliation{
  \institution{Carnegie Mellon University}
  \city{Pittsburgh, PA}
  \country{USA}
}
\renewcommand{\shortauthors}{Ferreira da Silva, et al.}

%% file: abstract.tex
\begin{abstract}
Scientific discovery is being revolutionized by AI and autonomous systems, yet current autonomous laboratories remain isolated islands unable to collaborate across institutions. We present the Autonomous Interconnected Science Lab Ecosystem (AISLE), a grassroots network transforming fragmented capabilities into a unified system that shorten the path from ideation to innovation to impact and accelerates discovery from decades to months. AISLE addresses five critical dimensions: (1)~cross-institutional equipment orchestration, (2)~intelligent data management with FAIR compliance, (3)~AI-agent driven orchestration grounded in scientific principles, (4)~interoperable agent communication interfaces, and (5)~AI/ML-integrated scientific education. By connecting autonomous agents across institutional boundaries, autonomous science can unlock research spaces inaccessible to traditional approaches while democratizing cutting-edge technologies. This paradigm shift toward collaborative autonomous science promises breakthroughs in sustainable energy, materials development, and public health.
\end{abstract}




%% file: sec_introduction.tex
\section{Introduction}

The scientific discovery process is undergoing a profound transformation, marked by the rise of automation, robotics, machine learning (ML), and artificial intelligence (AI). As we navigate the ``fourth industrial revolution"~\cite{xu2018fourth}, intelligent agents are emerging as the driving force behind a new paradigm where scientific exploration is no longer constrained by human cognitive limitations or decision-making timescales. The traditional research model, where human scientists manually design experiments, analyze data, and iterate hypotheses, is increasingly inadequate to address urgent global challenges in sustainable energy, climate science, materials development, and public health. Modern scientific instruments can generate data at rates that far outpace human analysis capabilities, creating a fundamental bottleneck in the discovery process. Autonomous science offers a transformative solution by combining AI, robotics, and computational workflows to accelerate discovery, eliminate human biases, and allow for the exploration of previously intractable research spaces~\cite{aww2024}. This mismatch between the length of human decision-making cycles and the potential speed of scientific exploration represents an opportunity for AI agent-driven workflows to revolutionize scientific practice~\cite{pauloski2025empowering}. 

Significant progress has been made toward the development of autonomous laboratories. Recent breakthroughs combining materials science and high-performance computing (HPC) have yielded tangible results in accelerating scientific discovery, evidenced by groundbreaking achievements such as the rapid discovery of novel metallic glasses through ML-enhanced high-throughput experimentation~\cite{ren2018accelerated}, successful isolation of gradient co-polymers using AI workflows in automated chemical synthesis~\cite{sumpter2023autonomous}, development of new organic semiconductor materials~\cite{seifrid2022autonomous} and electronic polymer films~\cite{wang2025autonomous}, and the accelerated discovery of materials for energy storage through AI-driven prediction~\cite{szymanski2023autonomous}. Initiatives like ORNL's INTERSECT~\cite{intersect}, ANL's Autonomous Research Laboratories~\cite{autonomous-discovery,gladier}, and PNNL's AT SCALE~\cite{atscale} showcase the potential of AI-driven autonomous systems. These approaches allow for faster, less expensive, and less labor-intensive research processes that ensure that data collection, synthesis, and analysis are conducted without human biases. With autonomous systems in place, the typical cycle of scientific problem solving, which often takes years or decades, can be shortened to months, weeks, or even days.

\begin{figure*}[!t]
    \centering
    \includegraphics[width=.99\linewidth]{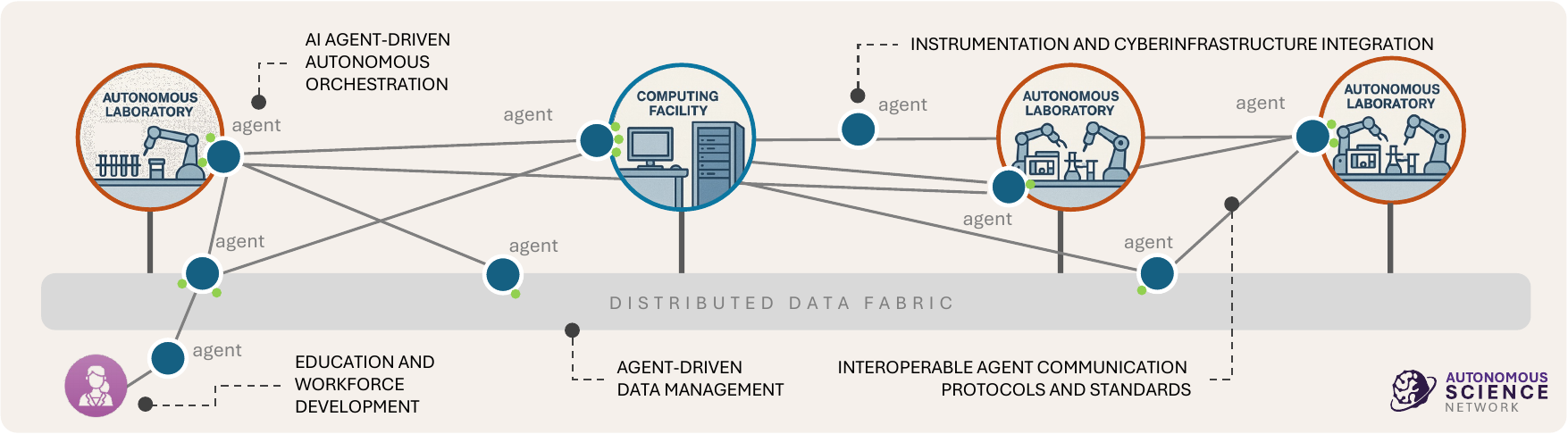}
    \vspace{-10pt}
    \caption{The AISLE network architecture illustrating the five critical dimensions for interconnected autonomous laboratories: Instruments and Cyberinfrastructure Integration, Agent-Driven Data Management, AI-Agent Driven Orchestration, Interoperable Agent Communication Protocols and Standards, and Education and Workforce Development, all connected through a distributed data fabric with intelligent agents.}
    \label{fig:overview}
\end{figure*}

Despite recent progress, the current landscape of autonomous science remains fragmented, with most systems operating in isolation and unable to communicate across institutions or disciplines. This contrasts with scientific workflows that naturally span multiple facilities, e.g., synthesizing a material in one lab, characterizing it at national user facilities, and running simulations on HPC systems. The federated nature of research infrastructures poses unique challenges: distributed resources have diverse access protocols, interactions between computational and experimental entities are asynchronous, and the dynamic availability of resources requires fault-tolerant and adaptive systems. Thus, we face critical challenges in developing a unified ecosystem, including: (1)~enabling communication between heterogeneous agent systems operating diverse scientific instruments; (2)~standardizing data and control interfaces to allow, for example, seamless agent collaboration; (3)~the development of AI/ML systems and AI agents that understand fundamental scientific principles; and (4)~creating adaptive, fault-tolerant agent coordination mechanisms that can navigate the complexities of distributed research infrastructure. 

A grassroots network approach is essential to connect autonomous capabilities across institutions, standardize protocols, democratize access, and accelerate the technology transition from research to application. This paper presents a vision for a nationwide grassroots network of interconnected autonomous laboratories (\textbf{AISLE}) that will transform scientific discovery. We discuss five critical dimensions for building this ecosystem (Fig.~\ref{fig:overview}):

\begin{enumerate}[nosep,leftmargin=2em,labelwidth=*,align=left]
    \item \emph{Instruments and Cyberinfrastructure Integration} that enables agents to orchestrate various scientific instruments across institutional boundaries.
    \item \emph{Agent-Driven Data Management} where autonomous agents actively curate, validate, and orchestrate scientific data across institutional boundaries while automatically enforcing the principles of FAIR~\cite{wilkinson2016fair}.
    \item \emph{AI Agent-Driven Autonomous Orchestration}, exploring robust hierarchical architectures that can leverage agentic capabilities of LLM-based agents to orchestrate traditional methods grounded in scientific knowledge and physics.
    \item \emph{Interoperable Agent Communication Interfaces and Standards} for multi-agent systems to enable seamlessly information exchange, activities coordination, and the integration of capabilities across institutional and disciplinary boundaries. 
    \item \emph{Education and Workforce Development}, preparing scientists for human-AI collaboration in environments increasingly dominated by autonomous systems while maintaining scientific rigor and critical thinking.
\end{enumerate}


%% file: sec_aisle.tex
\section{The AISLE Network}
\label{sec:aisle}

The Autonomous Interconnected Science Lab Ecosystem (AISLE, \url{https://autonomousscience.org}) represents a grassroots network dedicated to revolutionizing scientific discovery through interconnected autonomous laboratories. AISLE aims to dramatically accelerate the journey from ideation to innovation to market application, reducing time frames from decades to months, while enabling previously unattainable discoveries. This initiative integrates AI-ready hardware, software, and data infrastructure to create a nationwide capability that enhances the scope, speed, and responsiveness of scientific research. The core mission focuses on developing a cohesive cross-domain data fabric that optimizes workflows, enhances reproducibility, democratizes access to cutting-edge scientific technologies, and facilitates rapid technology transition across disciplines with direct applications to national priorities.

AISLE unites forward-thinking scientists, engineers, and technologists committed to reimagining scientific methodologies. This diverse community spans multiple disciplines including materials research, computer science, engineering, robotics, AI, accelerator science, chemistry, and data management. Rather than operating within formal institutional structures, AISLE functions as a collaborative knowledge exchange network where expertise flows across organizational boundaries, creating an environment where experimental insights inform computational approaches in a tightly coupled virtuous circle. The network's strength lies in its variety of perspectives, with members contributing complementary capabilities to build a foundation for transforming scientific experimentation through autonomous intelligent systems that augment human creativity while addressing society's most pressing challenges.

%% file: sec_dimensions.tex
\section{Critical Dimensions of the AISLE Network}
\label{sec:dimentions}

The realization of interconnected autonomous science laboratories requires a framework built upon five critical dimensions that collectively enable seamless collaboration between autonomous agents across institutional and disciplinary boundaries. These layers work synergistically to create an ecosystem greater than the sum of its parts. Each dimension presents unique challenges and research opportunities that must be addressed through coordinated development efforts, recognizing the interdependencies between technical infrastructure, intelligent systems, and human expertise.

\subsection{Instrument and CI Integration}
\label{sec:instrument}

Scientific instruments and cyberinfrastructure (CI) integration address how autonomous agents can orchestrate diverse experimental equipments and computational resources. Unlike traditional systems that function within institutional silos, interconnected autonomous laboratories require agents that can control instruments, manage data flows, and coordinate computational analyses across organizational boundaries. This integration is essential for accelerating materials discovery, where instruments such as electron microscopes, X-ray diffractometers, and synthesis robots generate heterogeneous data that must be processed through complex computational pipelines spanning multiple facilities.

\pp{Brief State-of-the-art}
Current integration approaches demonstrate promising directions in multiple scientific domains. The Materials Acceleration Platform (MAP) initiative exemplifies the international momentum towards fully automated laboratories, with several projects demonstrating end-to-end autonomous materials discovery workflows~\cite{stier2024materials}. The Academy middleware enables the deployment of federated agents on experimental and computational resources, providing abstractions to express stateful agents, and managing interagent coordination with experimental control~\cite{pauloski2025empowering}. These systems support asynchronous execution, heterogeneous resources, and high-throughput data flows essential for scientific computing. Practical communication frameworks are emerging, including popular ROS2 / DDS messaging protocols in robotics applications, and companion standards of OPC UA specifically designed for the integration of laboratory equipment~\cite{profanter2019opc}.  Self-driving labs like those described in the materials discovery domain have begun to integrate AI-ready hardware, software, and data infrastructure to create autonomous capabilities that optimize workflows and enhance reproducibility~\cite{madsci, decentralized-ai}. Physics-aware digital twins are increasingly being used for testing and validating autonomous workflows before deployment on physical instruments, reducing experimental risks and costs. Recent advancements in material- and process-efficient self-driving laboratories (SDLs) have demonstrated scalable and sustainable strategies for autonomous experimentation. These platforms combine miniaturized reaction vessels with AI-guided decision-making and multi-modal characterization to maximize information gained per experiment, minimize chemical waste, and significantly reduce operational costs. Notably, fluidic SDLs have achieved >100$\times$ data acquisition efficiency over traditional batch methods while maintaining reproducibility and closed-loop optimization capabilities~\cite{sadeghi2025self}. SDLs have proven particularly valuable in organic synthesis and semiconductor materials discovery, where multi-robot SDLs now coordinate synthesis, purification, and characterization steps through modular robotic orchestration. Integrating such SDLs into AISLE’s architecture can provide modular, high-throughput testbeds that are adaptable to distributed, cross-institutional networks. In chemistry, ChemCrow extends the capabilities of large language models (LLMs) by integrating 18 expert-designed tools for organic synthesis, drug discovery, and material design~\cite{m2024augmenting}. Biological research has seen the emergence of specialized systems, including CellAgent, which employs multiple expert agents (Planner, Executor, and Evaluator) to automate complex data analysis tasks~\cite{xiao2024cellagent}.

\pp{Challenges}
Integration faces significant technical and organizational challenges across domains. Scientific autonomous workflows must span various instruments ranging from established commercial products to custom-built research equipment not originally designed for networked automation~\cite{aww2024}. The heterogeneous nature of scientific instruments is addressed through concepts of ``cellular" or ``modular" laboratory design that standardize interfaces while allowing flexibility in equipment configuration. Recent implementations of autonomous science reveal additional systemic challenges. Software frameworks designed with these principles in mind have been recently introduced~\cite{madsci}. The automation of literature review remains a bottleneck, with frameworks that exhibit significant performance drops during the literature review phases compared to other research stages~\cite{gridach2025agentic}. Trustworthiness and reliability concerns pose obstacles, as current systems struggle to avoid overfitting and maintain predictable behavior in diverse scientific contexts. Furthermore, coordination challenges in multi-agent systems become amplified in distributed experimental environments, where communication failures can lead to resource conflicts, protocol deviations, or safety hazards. Finally, critical organizational barriers include intellectual property management and liability concerns when cross-institutional failures occur, which are often overlooked but will significantly constrain real-world deployments.

\pp{Research Priorities}
Initial efforts should focus on developing domain-specific integration frameworks for common scientific instruments and establishing reference implementations that demonstrate cross-facility workflows in targeted domains such as materials science or accelerator physics. These foundational activities include expanding currently available interfaces to support a broader range of experimental equipment and implementing basic orchestration protocols for distributed instrument control. Building upon these capabilities, more sophisticated infrastructure development involves creating standardized hardware abstraction layers and robust security models for multi-institutional access through enhanced collaboration with instrument vendors to develop developer-friendly interfaces. Advanced implementation phases require establishing adaptive fault-tolerant coordination mechanisms that can handle complex resource dependencies and dynamic network conditions, along with governance frameworks that maintain institutional autonomy while enabling seamless collaboration. Future research needs include developing self-describing instruments with semantic descriptors for capabilities, automated calibration protocols that enable instruments to ``plug in" without manual setup, and robust human-in-the-loop safeguards that allow operators to override %
\milestones{
\begin{itemize}[nosep,leftmargin=.5em,labelwidth=*,align=left]
    \item[\bf\footnotesize M1.] Establish common integration interfaces for scientific instruments with vendor-agnostic hardware abstraction layers and API development via an Instrument API Consortium.

    \item[\bf\footnotesize M2.] Demonstrate end-to-end autonomous workflows across institutions for seamless experimental and computational resource orchestration through secure multi-domain cyberinfrastructure networks.

    \item[\bf\footnotesize M3.] Deploy federated cyberinfrastructure with standardized frameworks, fault-tolerant coordination mechanisms, and adaptive resource management with zero-trust security and physics-aware digital twins for workflow validation.

    \item[\bf\footnotesize M4.] Scalable national framework supporting heterogeneous instruments for near real-time data flows with self-describing instruments, automated calibration, and human-in-the-loop override capabilities.
\end{itemize}
}%
\noindent autonomous agents sending laboratory robots out-of-specification commands. The most transformational aspects involve the deployment of self-configuring cyberinfrastructure that can automatically adapt to new instrument types and evolving network topologies, ultimately creating a resilient ecosystem capable of supporting autonomous experimental workflows across diverse scientific domains and institutional boundaries.

\subsection{Agent-Driven Data Management}
\label{sec:data}

Agent-driven data management represents a paradigm shift from traditional centralized data repositories to intelligent distributed systems where autonomous agents actively curate, validate, and orchestrate scientific data across institutional boundaries. Data management agents act as intelligent intermediaries that understand scientific context, enforce FAIR principles in near real time, and make data AI-ready at the source by curating, annotating, and organizing it as it is being collected across heterogeneous experimental facilities. These agents must handle the full data lifecycle, from real-time capture during autonomous experiments to long-term preservation and cross-institutional sharing, while maintaining data quality, provenance, and compliance with diverse institutional policies. Unlike passive data storage systems, agent-driven approaches actively monitor data streams, perform intelligent quality assessment, and facilitate dynamic data federation that adapts to the evolving needs of autonomous workflows.

\pp{Brief State-of-the-art}
Current data management approaches in scientific workflows rely mainly on centralized repositories and domain-specific standards, with successful examples including Materials Commons and Protein Data Bank (PDB)~\cite{velankar2021protein}. The principles of FAIR (Findable, Accessible, Interoperable, Reusable) data have gained widespread adoption as a framework for scientific data management, although implementation remains inconsistent between domains and institutions~\cite{wilkinson2016fair}. FAIR AI models and emerging FAIR data meshes demonstrate advanced approaches to federated scientific data management~\cite{ravi2022fair}. Globus employs cloud-hosted management logic to coordinate activities across thousands of storage and computing systems worldwide~\cite{chard2023globus}. 
The National Science Data Fabric (NSDF)~\cite{luettgau2023nsdf} and the National Data Platform (NDP)~\cite{parashar2023toward} aim to develop a federated approach to data management that coordinates networking, storage, and computing services through distributed entry points. Recent advances have demonstrated the potential of AI-driven metadata extraction and automated data curation systems that can intelligently interpret experimental contexts and integrate external information sources. Systems like ProxyStore enable efficient data transfer through pass-by-reference semantics in distributed computing environments, allowing large datasets to be shared without duplicating storage~\cite{pauloski2024accelerating}.

\pp{Challenges}
Agent-driven autonomous laboratories face fundamental data management challenges that extend far beyond traditional workflows. At the technical level, the variety of data formats and file structures generated by different instruments makes it difficult to create vendor-agnostic abstract data interfaces that can support autonomous workflows across institutional boundaries. This technical heterogeneity consists of operational variations, as experimental protocols vary between institutions, environmental conditions affect reproducibility, and equipment calibration differences introduce systematic variations that current systems cannot automatically reconcile. A critical research gap involves dynamic schema evolution: how autonomous agents can negotiate schema changes when encountering new experiment types without manual intervention. Beyond heterogeneity, autonomous systems must also deal with unprecedented data volumes. Near real-time data streams from modern instruments generate volumes that exceed human processing capabilities, requiring intelligent filtering and prioritization mechanisms that can distinguish between routine measurements and anomalous conditions requiring immediate attention. This volume challenge is closely linked to data quality concerns, as ``bad" or ``imbalanced" data can propagate through AI-driven decision chains, potentially compromising entire experimental campaigns. Unlike traditional approaches that treat all data equally, autonomous systems require qualification mechanisms that can automatically assess data reliability based on experimental conditions, instrument status, and historical patterns. Privacy and regulatory constraints (e.g., HIPAA compliance for biological laboratories) create additional barriers that complicate federated data sharing across institutional boundaries. These challenges are further amplified by the distributed nature of autonomous laboratories, which creates complex requirements to maintain data provenance and ensure the traceability of decisions made by AI agents on multiple facilities and time scales.

\pp{Research Priorities}
Developing agent-driven data management requires establishing adaptive, domain-agnostic frameworks that can evolve with scientific advances while maintaining interoperability across diverse research environments. Priority should be given to implementing data mesh architectures in which each laboratory maintains a federated node with standardized interfaces, complemented by global discovery indices~\cite{bergamasco2012knowledge}. Rather than enforcing strict standardization, data schemas should support both explicit and implicit structures to enable seamless integration of heterogeneous scientific instruments and computing systems while preserving institutional autonomy and data sovereignty. AI agents can leverage implicit data schemas by inferring structure and extracting useful information directly from diverse data sources, formats, and contexts. While early adoption of interoperable frameworks such as HDF5 or JSON-LD can provide a useful foundation, advanced AI systems for metadata collection must be able to interpret complex scientific contexts, extract insights from laboratory notebooks and equipment logs, and incorporate environmental data without relying solely on predefined annotation. Integration of data provenance frameworks (e.g., PROV-O~\cite{lebo2013prov}) into instrument middleware will ensure comprehensive traceability of autonomous decisions across %
\milestones{
\begin{itemize}[nosep,leftmargin=.5em,labelwidth=*,align=left]
    \item[\bf\footnotesize M5.] Develop AI-driven metadata systems with automated annotation of experimental data in multiple domains, achieving high accuracy without human intervention.
    
    \item[\bf\footnotesize M6.] Deploy federated data mesh architecture with common APIs, cross-institutional discovery capabilities, and autonomous FAIR data governance.
    
    \item[\bf\footnotesize M7.] Implement near real-time data processing infrastructure supporting high-velocity scientific streams with automated quality assessment, provenance tracking, and regulatory compliance frameworks.    
\end{itemize}
}
\noindent  distributed facilities. Federated data management architectures must be designed that enable cross-institutional collaboration while respecting privacy constraints, intellectual property rights, and regulatory compliance requirements. Near real-time data processing pipelines should be developed that can handle high-velocity scientific data streams, perform intelligent data reduction and compression, and trigger appropriate responses to critical experimental conditions. Community-driven approaches, including data annotation sprints, should be promoted to accelerate the development of high-quality training datasets for autonomous systems.

\subsection{AI Agent-Driven Autonomous Orchestration}

AI-driven autonomous orchestration represents the cognitive core of interconnected autonomous laboratories, where intelligent agents must navigate complex scientific decision spaces while maintaining alignment with fundamental scientific principles. Modern LLM-based agents emerge as orchestrators coordinating specialized techniques: Gaussian processes for uncertainty quantification, Bayesian optimization for sample efficiency, and reinforcement learning for dynamic control, enabled by natural language understanding of scientific goals. This autonomy must be deployed as composable building blocks in the scientific ecosystem, recognizing both the capabilities and limitations of the underlying models while integrating verification infrastructure throughout autonomous workflows. These agents leverage instruments (Section~\ref{sec:instrument}) as actuators for experimental execution and synthesize real-time data streams (Section~\ref{sec:data}) with literature and cross-facility insights to enable trustworthy autonomous discovery.

\pp{Brief State-of-the-art} Current AI-driven capabilities in autonomous science demonstrate significant progress across multiple domains. Recent comprehensive surveys highlight the integration of robotics, AI, and automation in sustainable chemistry applications, demonstrating the maturation of autonomous laboratory technologies~\cite{sadeghi2024engineering}. Recent advances include hybrid AI architectures that combine data-driven learning with fundamental physical and chemical principles~\cite{al2024autonomous}, and human-autonomy teaming frameworks with adaptive trust calibration systems~\cite{hagos2024ai}. Specific AI techniques showing promise include Gaussian processes for sample-efficient Bayesian optimization in materials discovery, reinforcement learning for dynamic experimental scheduling, and active transfer learning approaches enabling knowledge sharing between laboratories. Notable examples include Smart Dope, which navigates $10^{13}$ possible synthesis conditions to discover optimal quantum dot formulations~\cite{sadeghi2024engineering}. In nanomaterial synthesis and homogeneous catalysis, autonomous frameworks leverage nested discrete-continuous Bayesian optimization strategies that reflect real-world experimental constraints~\cite{sadeghi2025self}. These approaches improve optimization efficiency by structuring search spaces to reflect hardware constraints, which have been successfully applied in reaction condition optimization tasks. In recent years, the emergence of large language model (LLM) based agents as general-purpose scientific actors impose new opportunities in autonomous science. Such applications of foundation models show natural language understanding of scientific goals and the ability to orchestrate multiple specialized AI tools, as demonstrated by efforts such as the LLNL HPC-LLM agents~\cite{hpc-llm2025} and DOE autonomous discovery initiatives~\cite{autonomous-discovery}. Built on top of LLMs, augmenting their input context with various tools interacting with the world, these agents show emergent capabilities that include hypothesis generation, cross-domain knowledge transfer, adaptive experimental strategies, and high-level orchestration that can fundamentally transform scientific discovery.

\pp{Challenges}
With the newly found opportunities in LLM-based AI agents, autonomous decision making in scientific contexts faces fundamental challenges in integration, orchestration, reliability, and grounding. LLM agents require careful investigation of their capability, limits, and requirements, identifying their place in the scientific ecosystem and infrastructure. They are probabilistic in nature, higher-latency, and resource intensive compared to traditional methods, and are difficult to verify. These agents are part of the ecosystem as an orchestrator rather than a replacement for existing techniques. Though, challenges lie in creating a robust, well-integrated architecture that can support seamless transition between techniques and stages potentially spread across scientific domains, context, geographic locations, and long time horizons. Also, while capable, the probabilistic nature is a key challenge in integrating LLM-based AI agents, especially in an ecosystem attuned to determinism. It is unclear how one would guarantee reproducible scientific outcomes with this new non-determinism in effect. Further, there are no guarantees whether the solutions driven by these systems would be grounded in scientific knowledge and physics.

\pp{Research Priorities}
Three interconnected research thrusts are key for advancing agent-driven autonomous orchestration in scientific contexts: (1)~design hierarchical architectures and infrastructure for efficient orchestration, deploying AI agents in the ecosystem of scientific methods abstracted as actuators, coordinating tasks depending on the model capabilities and system requirements (e.g., compute usage, latency); (2)~infrastructure for verification and validation for AI agents incorporating digital twin-based in-situ simulations, formal methods, symbolic verification methods to enforce logical, physics-based constraints as hard boundaries when agents work towards optimal solutions or discovery; and (3)~distributed, real-time knowledge integration that helps the operations of AI agents grounded to scientific knowledge beyond static information retrieval or fine-tuning of models, especially considering scientific campaigns distributed across facilities, instruments, and many teams coordinating distributed, asynchronous, real-time evolution of knowledge. Priority should be compositional scientific AI systems that focus on composing LLM reasoning with traditional ML methods, verification tools, and dynamic knowledge bases.
\milestones{
\begin{itemize}[nosep,leftmargin=.5em,labelwidth=*,align=left]
    \item[\bf\footnotesize M8.] Demonstrate hierarchical architectures which LLM agents orchestrate traditional methods through domain-specific scientific interfaces, achieving 3x speedup over manual orchestration and >95\% experimental correctness versus agent usage without verification tools.

    \item[\bf\footnotesize M9.] Deploy a knowledge integration system with 3+ facilities, propagating insights across sites in real-time to reduce required experiments by >30\% while achieving >90\% scientist approval of reasoning traces.
\end{itemize}
}

\subsection{Interoperable Agent Communication}

Interoperable agent communication interfaces and standards form the foundational infrastructure that enables diverse autonomous systems to seamlessly exchange information, coordinate activities, and integrate capabilities across institutional and disciplinary boundaries. Unlike traditional point-to-point communication approaches, autonomous scientific laboratories require sophisticated agent communication frameworks that can handle asynchronous interactions, manage complex state dependencies, and maintain coherent coordination across distributed experimental and computational resources. These interfaces must support various interaction patterns including peer-to-peer agent coordination, hierarchical command structures, and emergent collaborative behaviors while ensuring reliability, security, and fault tolerance in multi-institutional research environments.

\pp{Brief State-of-the-art}
Current agent communication in autonomous science is based primarily on domain-specific solutions and proprietary interfaces, with limited standardization between platforms. Several orchestration architectures have emerged, including ChemOS 2.0 for coordinating communication and data exchange among chemical synthesis instruments~\cite{sim2024chemos}, Globus automation services~\cite{gladier}, and the Academy middleware, which enables the deployment of federated agents across experimental and computational resources while managing inter-agent coordination~\cite{pauloski2024accelerating}. Modern implementations increasingly leverage containerization technologies, with agent microservices communicating through high-performance protocols such as gRPC for synchronous operations and AMQP for asynchronous message queueing in distributed workflows. Human-autonomy teaming frameworks have demonstrated the importance of bidirectional communication approaches that transform automation from tools to collaborative teammates, enabling dynamic information exchange and joint decision-making processes~\cite{hagos2024ai}. The INTERSECT initiative developed a federated architecture to coordinate autonomous processes across distributed scientific infrastructure~\cite{mintz2023towards}.  Recent advances in large language models have shown the potential for natural language interfaces that could serve as universal communication bridges between human operators and diverse autonomous systems~\cite{hysmith2024future}.

\pp{Challenges}
The heterogeneous nature of scientific instruments creates fundamental complications in the development of standardized interfaces, particularly when instruments use proprietary control systems with limited API access or vendor-specific protocols. Multi-domain network architectures with complex firewalls and access controls further complicate efforts to orchestrate distributed experiments, as autonomous agents must navigate diverse institutional security policies, authentication systems, and network topologies while maintaining secure and reliable communication channels. Zero-trust network architectures present additional complexity, requiring continuous authentication and authorization of agent interactions while maintaining low-latency communication necessary for near real-time experimental control. The asynchronous nature of scientific workflows introduces additional complexity, as agents must coordinate across different timescales, from near real-time instrument control that requires millisecond responses to long-term experimental campaigns that span weeks or months, while managing state consistency and gracefully handling communication failures. Semantic interoperability presents ongoing challenges, as different scientific domains use varying data formats, measurement units, and conceptual frameworks that must be harmonized for effective cross-disciplinary agent collaboration. The distributed and federated nature of autonomous laboratory networks creates scalability concerns, as communication protocols must efficiently handle coordination among potentially hundreds of agents while avoiding bottlenecks and ensuring fault tolerance when individual nodes or communication links fail.
 
\pp{Research Priorities}
Robust interoperable communication requires developing layered protocol architectures that separate concerns across physical networking, message formatting, semantic interpretation, and coordination logic levels. Priority should be given to creating vendor-agnostic hardware abstraction layers with standardized APIs that can interface with instruments from multiple manufacturers while providing consistent communication interfaces for autonomous agents. Advanced message-oriented middleware solutions must be developed that support asynchronous communication patterns, reliable message delivery, and automatic failover mechanisms essential for distributed scientific workflows operating across institutional boundaries. Semantic interoperability frameworks should incorporate domain ontologies and knowledge graphs that enable agents to automatically translate between different scientific vocabularies and data representations while preserving meaning and context. Security and authentication protocols specifically designed for multi-institutional scientific collaboration must balance access control requirements with the need for seamless agent interaction, incorporating technologies such as federated identity management and attribute-based access control. Finally, self-organizing communication protocols should be explored that allow agent networks to automatically discover capabilities, negotiate communication parameters, and adapt to changing network topologies without requiring centralized configuration management. Testbed environments should demonstrate autonomous agent ecosystems where containerized services can dynamically discover and interact with heterogeneous scientific instruments through standardized protocols, validating both technical performance and security frameworks under realistic multi-institutional conditions.
\milestones{
\begin{itemize}[nosep,leftmargin=.5em,labelwidth=*,align=left]
    \item[\bf\footnotesize M10.] Deploy containerized agent microservices with standardized gRPC/AMQP communication protocols across multiple DOE laboratory facilities, demonstrating cross-vendor instrument control and federated identity integration.

    \item[\bf\footnotesize M11.] Develop zero-trust communication infrastructure supporting autonomous agent coordination with sub-second latency, automatic failover, and continuous authentication across institutional boundaries.

    \item[\bf\footnotesize M12.] Demonstrate self-discovering agent networks using DNS-SD and distributed service registries, enabling dynamic reconfiguration and capability negotiation in geographically distributed research facilities.
\end{itemize}
}

\subsection{Education and Workforce Development}

The successful deployment of interconnected autonomous science laboratories requires fundamental transformations in scientific education to prepare researchers for environments increasingly dominated by AI-driven systems. Unlike traditional education focused on domain-specific knowledge and manual techniques, autonomous science demands interdisciplinary competencies spanning AI/ML methods, computational and workflow thinking, human-machine collaboration, and ethical reasoning. Educational programs must evolve to prepare scientists who can collaborate effectively with autonomous agents, understand AI decision-making processes, and maintain scientific rigor while leveraging computational tools that augment human creativity.

\pp{Brief State-of-the-art}
Current scientific education largely treats AI/ML as supplementary rather than integral to scientific methodology, leading to competency gaps in preparing researchers for autonomous laboratory environments~\cite{aww2024}. While national initiatives such as the NSF's AI Institutes and DOE's SciDAC programs are beginning to address these gaps through dedicated education pillars, a broader and more integrated approach is needed. Effective workforce development for autonomous science must encompass not only AI/ML but also robotics, software engineering, networking, and laboratory safety. Emerging cross-disciplinary training programs, virtual lab environments, and AI-enhanced computational tools offer promising foundations. Human-autonomy teaming frameworks from operational domains emphasize the importance of understanding AI capabilities and limitations while maintaining human oversight~\cite{hagos2024ai}. However, comprehensive curriculum redesign remains limited across institutions, and must evolve to include hands-on, experiential learning that mirrors the complexity and interdisciplinarity of real-world autonomous research settings.

\pp{Challenges}
Fundamental challenges include balancing automation capabilities with core scientific understanding, as excessive AI/ML reliance risks creating scientists lacking foundational knowledge to critically evaluate automated results. A critical assessment gap has emerged: current evaluation methods cannot effectively measure students' ability to ``collaborate with AI," requiring new frameworks adapted from fields such as medical simulation training where human-technology interaction is rigorously assessed. Rapid advancement in AI/ML creates challenges in curriculum development, while faculty development presents barriers as many educators lack the necessary AI/ML expertise. The interdisciplinary nature of autonomous science research requires institutional restructuring across traditional departmental boundaries, and hands-on training presents logistical obstacles due to limited access to sophisticated autonomous laboratory infrastructures. Furthermore, traditional assessment methods do not adequately capture human-AI collaboration competencies, and ensuring equitable access becomes critical to preventing workforce disparities.

\pp{Research Priorities}
Curriculum redesign must integrate AI/ML competencies with fundamental scientific principles through active learning and authentic research experiences that demonstrate human-machine synergy. Priority areas include creating modular educational frameworks adaptable across disciplines, developing virtual training environments for immersive autonomous laboratory experiences, and establishing faculty development programs that maintain the emphasis on critical thinking and scientific reasoning. Industry-academic partnerships should provide authentic experiences, while assessment methodologies must assess human-AI collaboration competencies including AI decision interpretation and appropriate trust calibration. Ethical reasoning frameworks must be integrated throughout, ensuring that future scientists understand the societal implications of AI-driven research.

\milestones{
\begin{itemize}[nosep,leftmargin=.5em,labelwidth=*,align=left]
    \item[\bf\footnotesize M13.] Launch a national autonomous science education consortium that integrates NSF AI Institutes and DOE SciDAC programs, with standardized autonomous laboratory collaboration curricula.
    
    \item[\bf\footnotesize M14.] Deploy educational infrastructure including immersive virtual laboratory environments that simulate autonomous systems in multiple scientific domains, industry-academic partnership programs, and assessment methodologies for human-AI collaboration competencies with measurable learning outcomes.
\end{itemize}
}

%% file: sec_conclusion.tex
\section{Conclusion}
\label{sec:conclusion}

The AISLE network represents a transformative vision for scientific discovery, where interconnected autonomous laboratories transcend institutional and disciplinary boundaries to create a unified system capable of accelerating breakthroughs from decades to months. By addressing the five critical dimensions of instruments integration, agent-driven data management, AI-agent driven orchestration, interoperable communication protocols, and workforce development, AISLE will unlock research spaces previously inaccessible to traditional human-centered approaches while democratizing access to cutting-edge scientific technologies. The successful implementation of this grassroots network promises revolutionary advances in science through collaborative autonomous agents that augment human creativity and scientific insight. AISLE is uniquely positioned to catalyze progress across national initiatives such as the Materials Genome Initiative and the CHIPS and Science Act by enabling testbeds for co-design of materials and devices via autonomous experimentation. The federation of domain-specific SDLs through AISLE’s proposed agent fabric offers a practical blueprint for large-scale coordination of AI, robotics, and data infrastructures across the U.S. science enterprise. Importantly, these efforts should also prioritize inclusion of resource-constrained institutions by supporting portable, low-footprint SDL modules that contribute to the broader network while enabling equitable access to advanced automation technologies. Future work will focus on establishing pilot testbeds that demonstrate cross-institutional autonomous workflows, developing standardized protocols for multi-vendor instrument integration, and creating educational frameworks that prepare the next generation of scientists for this new paradigm of AI-augmented discovery.